# Software, Attacker and Asset-centric Approach for Improving Security in System Development Process

1st Abdul Hadi bin Abdul Rahman
*School of Computer Science & Engineering,*
*Taylor's University*
Selangor, Malaysia
abdulhadi96.ar@gmail.com

2nd Abdullah Nazir
*School of Computer Science & Engineering,*
*Taylor's University*
Selangor, Malaysia
abdullah97.nazir@gmail.com

3rd Kim Tae Hyun
*School of Computer Science & Engineering,*
*Taylor's University*
Selangor, Malaysia
kimth0312@gmail.com

4th Tan Horng Yarng
*School of Computer Science & Engineering,*
*Taylor's University*
Selangor, Malaysia
ethan990720@gmail.com

5th Fatima-tuz-Zahra
*School of Computer Science & Engineering,*
*Taylor's University*
Selangor, Malaysia
fatemah.tuz.zahra@gmail.com

**Abstract –** Secure development process is a procedure taken by developers to ensure the programs developed are following the general security standards and will always be up to date so that the outcomes are well secured and obedient. As a software developer, it is very crucial to implement and develop a highly secured and reliable program for clients and users. In this current digital world where everything is advancing faster than we can ever think of, most of the old security policies can no longer be implemented alone. The consequences and impacts that could be brought upon a company are really huge if the software applications are not secured according to the modern trend. Therefore, in this paper research is done to asses the security integration in software development process. The concept and the purpose of this research is to provide insight about the current issues and challenges faced by most of the software developers in terms of secure software development. With a better and clearer explanation of these issues, challenges, and methodologies adopted to overcome them are discussed which can potentially provide a better and higher level of security along with better software programmers and client relationship. To comply with future demands and threats, security concerns need to be involved in all phases while developing a software system. Therefore, an effort is made to investigate and contribute to this domain through this paper.

## 1 Introduction

Currently, every single individual is having a convenient lifestyle and it is way different compared to how our ancestors were living back in the time. With that being said, most of the people around the world will have at least a technology device that can allow them to be connected to the internet such as a mobile phone, tablet, a personal computer and more nowadays. All of the technology devices required a running and secured software system to be compatible with the hardware itself so that the device itself can run smoothly without any errors and flaws. Not only that, but it also gives users to be worry-free when they are using those devices. Users are always worried about their data's safety in an environment where most of the functions have been digitized [1] and automated [2]. They must think twice before they perform certain actions like uploading their personal and sensitive information such as photos, addresses, credit card information and personal details into cloud service providers like Google Cloud, OneDrive and iCloud. If the software system is not properly designed and secured, users will be at risk where their data and information or even devices can get compromised and hacked by the cybercriminals. With these data and information being compromised by the attackers, they have an unauthorized access into it and they can use those data to demand a ransom or even sell those data in the black market for them to generate an income as a hacker [3]. Even if a software system is proved to be fully secured from the developer side, it can still be vulnerable to the attackers as there are the possibilities of users' negligence and also the network of the device connected to is not secured [3]. In order to strengthen the security of a software system, developers must implement a policy which is a phase-by-phase process known as Secure Software Development Lifecycle (SDLC). There are a few Secure SDLC models such as Agile Methodology, Iterative Methodology and Waterfall Methodology [4].

All of these models consist of common phases like Planning Phase, Design Phase, Implementation Phase, Test Planning Phase, Testing Phase and Development and Maintenance Phase [4]. With the assistance of Secure SDLC, developers can easily reduce flaws, weaknesses and vulnerabilities at the early phases of the software development stage. This can also help decrease the risks of the stakeholder from having a huge loss due to the faulty software system [5]. It is due to the reason that if a software system is detected to have weak points for intruders to sneak into the system database, it could cause a huge loss whether to the stakeholders or even to the software company that software developers are working in. Particularly, if the software or application is web-based, it means it can become vulnerable to internet-based security attacks if security features are not implemented in various phases of software development. For example, attackers may use botnets to trigger denial of service attacks [6] or ransomware attacks [7]. Not only that, secure SDLC also helps the project team to improve the quality of the product outcomes. By segregating the whole project into phases, the secure software developer not only takes the last phase seriously but the whole lifecycle phases. Also, a higher quality and a reassured software will be developed as each of the phases includes actions taken like security evaluation to ensure the project is flawless, and all of the project team members will live an easier life without having to worry and stressed-out what issues might be found if the security evaluation is done at the very last step before the project rolled out to market and do they have sufficient time to fix it before the due date [8].

Once the Secure SDLC is implemented into the project, the secure software developer can basically say that the software system is fully functional and fully protected from all kinds of current cybersecurity attacks. Though it was fully protected from the current cybersecurity attacks, software developers are still necessary to think ahead at all times and maintain the security and safety of their projects as there are always spaces for everyone to improve their skills and we cannot easily let our guard down and allow the attackers to have the opportunity to attack the system we built. In this paper, an in-depth insight of a secure development process for the developers, challenges faced, and approaches used to conquer it. Furthermore, this report also consists of our own analysis on the given topic in order to debut a research report that readers are able to analyse in an unbiased point of view.

## 2 Literature Review

### 2.1 GQM Approach and Secure SDLC

As information systems are becoming more important nowadays, it is obvious that the number of information security threats are increasing. Due it has been quite a long time since the software security issues appeared and software have more active roles in the modern era, we might assume that the process for building secure software has also evolved because information systems are to be protected against threats. However, for some reasons, the process of software development has not changed as we expected, and it did not include an exclusive phase for security [9]. Although there are various software development methodologies depending on the type, size and available resources as well as systematic engineering processes such as SDLC (Software Development Life Cycle), the number of security issues have still not dropped.

Many researchers have reasoned that one of the most important reasons is that when developers build software, they do not include security in any phase of the SDLC even though the experts have claimed solutions for this problem since the 1980s. The basic concept of their proposal is to combine the security with each phase of the process [10]. One experience from major software developers has shown that having security in each phase of the process helps to cut down vulnerabilities by 80% on average [11]. So why is the number of security threats increasing instead of dropping? The answer is simple - Software companies do not pay attention to security. However, it is also evident that not only the companies are to blame. Deploying security features in the process phase is not an easy task and therefore they need references to follow. At the same time, however, it is hardly possible to find any guidance documents or recommendations, which are meant to be used to conduct testing. There are the most widely used documents for software certification, but its application is pretty strict. Thus, it is urgent to have the development of procedural guidelines for secure software development. Fig. 1 shows various security components that can be included in each phase of software development lifecycle.

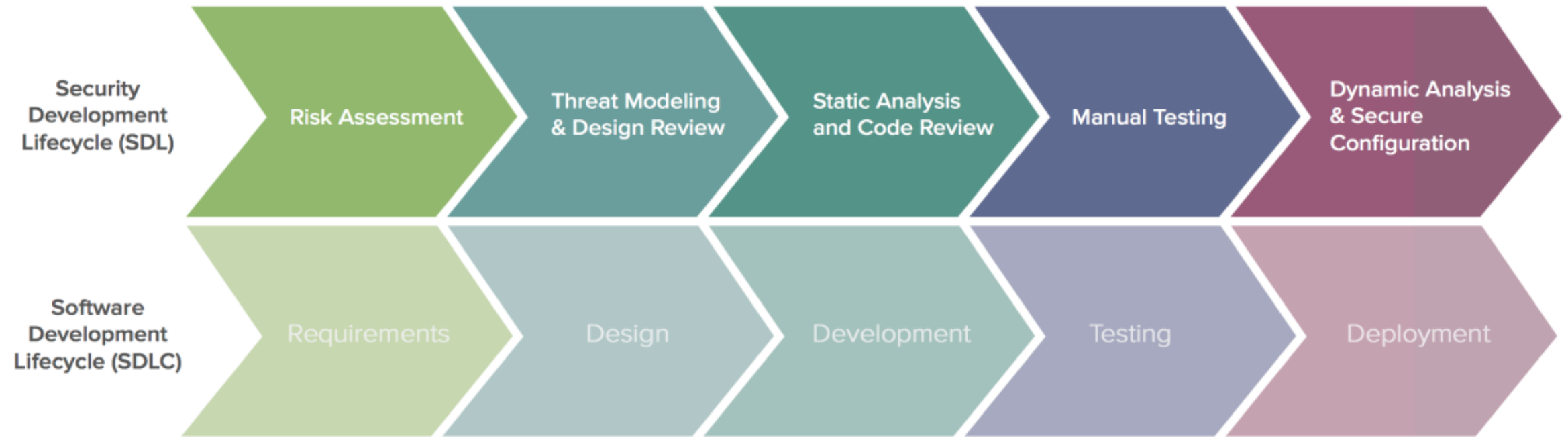

Fig. 1. A traditional SDLC versus a secure SDLC [12].

Before we go into how to build secure software, there are three principals we need to know. These principles are to attain security in the computer system - confidentiality, integrity and availability. These are also referred to as CIA. Confidentiality sets a set of rules and restrictions which help limit access to confidential information and make users' data secret and private [13]. This is needed to protect data from unauthorized viewers. Integrity ensures the whole data structure to be integrated as a fundamental concept of information security [14]. In simple words, the data or objects should not be changed by unauthorized entities. Availability means that service should be available to all users without any malfunction. Availability is guaranteed the most when hardware is maintained well [15]. Keeping these three principals in mind, let us talk about how we are going to measure the security. It seems hardly able to measure it since it is invisible, and we have no clue where we should give the highest and lowest score. In order to solve this problem, we are deploying security metrics called Goal-Question-Metric (GQM) approach which is the most widely used approach to develop security metrics and assess the security risks. When we define security metrics, we set the metrics as a standard term as security level, security indicators and security performance [16].

To briefly discuss the GQM approach, the Goal-Question-Metric approach is based on the assumption that when we measure something, we have to specify goals first and match those goals to the data, and lastly give a framework that interprets the data to the goals. GQM can be described on three levels - Conceptual level (Goal), Operational level (Question) and Quantitative level (Metric) [17]. Fig. 2 shows the hierarchical GQM approach.

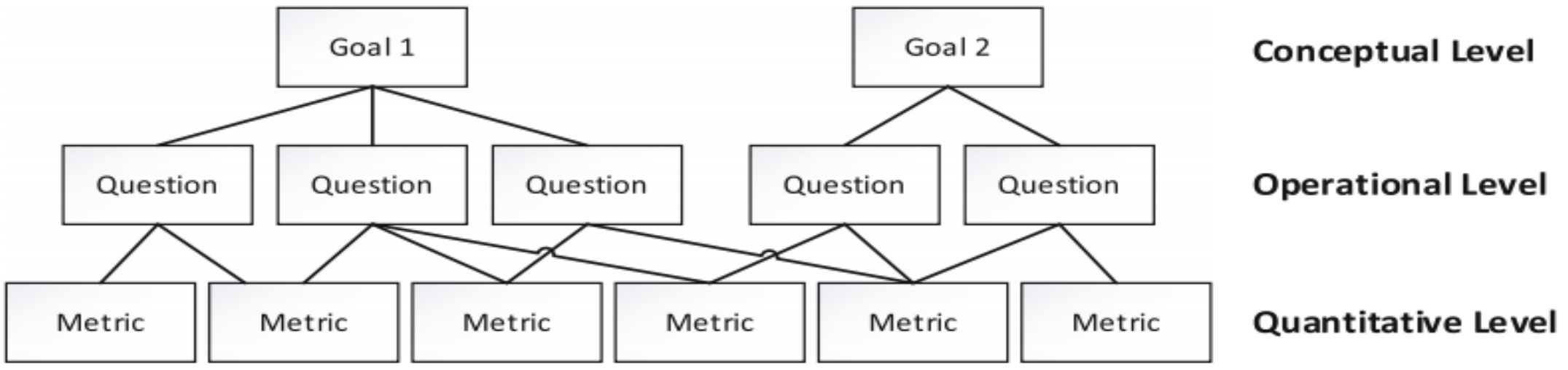

Fig. 2. GQM hierarchical approach [17][18].

In Fig. 2 it is shown that the goals are categorized in the conceptual level. In the conceptual level, goals are defined for objects and this is the most important part in GQM approach. Goals are to be described clearly and appropriately [19]. Next, we have questions at the operational level. A set of questions is used to define models of the object to describe the evaluation or accomplishment of a particular goal. Lastly, we have metrics where we evaluate the security and answer the questions. With the given information, we are now going to build a process in developing secure software, where there are 7 phases in total - Pre-Requirement Phase, Requirement Gathering and Analysis Phase, Software Design Phase, Coding Phase, Implementation Phase, Testing Phase and Maintenance Phase. You may think goals specified in figure 2 are phases and questions are activities we are going to specify [20].

First of all, we have a Pre-Requirement Phase. Pre-Requirement Phase should be stated before developers actually dive into building software. This phase must be elaborated deliberately since it contains the foundation for all the activities beginning from the early stage to the final stage. Mostly this phase is conducted by senior developers with stakeholders. Following questions are posed and answered in this phase: 1) Does the software really need the security feature, 2) is the security feature deployable, 3) what security features need to be deployed if required, 4) how will these features be implemented in the software. Next phase is the Requirement Gathering and Analysis Phase. In this phase, the developers should pay attention to estimate it properly. This phase gives a clearer view of the whole picture of software. In order to prevent any issue during building software, this phase has to be conducted thoroughly. This also helps the company to finalize the deadline for the work. Following are the questions specific to this phase: 1) How many security features are needed?, 2) what are the top five most required features, 3) which security aspects can be skipped, 4) what are the most time-costly security features.

The third phase is the Software Design Phase. This phase is just prior to the Coding Phase, which means all plans should be successfully made by this phase. We are to consider the following items during the design phase: 1) a basic outline for every security feature, 2) pointing out the module which is sensitive to software security, 3) relationship between dependencies and possible vulnerabilities. The fourth phase is the Coding Phase. When the software design is structured out, developers will then initiate by writing the program code with the selected programming language. This phase will take the longest time for it to complete and there are certain concerns that developers should take note of are: 1) choice of programming language, 2) alignment of program's purpose with the client's requirements, 3) updating the client about the progress, 4) coding experience of programmers employed for the job.

The fifth phase is the implementation stage where the software is actually made. In this phase the requirements given by the client are used to construct the application itself [21]. Security activities in this phase are based around ensuring safe practices are used to construct the application, dynamic testing is done, and the function of the application is tested for vulnerabilities [22]. Many automated tools can be used to test the application. Things to consider are: 1) checking whether the secure practices have been used to build the application, 2) check for vulnerabilities in any feature of the application, 3) performance results of the automated testing tools. The sixth phase of Secure SDLC is the Testing Phase. This phase is where the project team has done the source code and will run an application testing to benchmark and see whether the output is aligned with clients' requirements or not. This phase will be complete until there are no flaws, reliable and can fulfil all the requirements stated. The things that developers need to consider are: 1) testing type required, 2) testing environment compatibility with the software application, 3) time required to conduct the testing as well time required for fixing any issues, 4) number of cases that need to go through security testing process.

The seventh and final phase is the deployment and maintenance phase which is done after the testing phase and the software is ready for the end client. In the deployment phase the software is put into production and delivered to the client. The maintenance phase ensures the software keeps functioning as it is intended to. It handles errors if they occur after the deployment of the software and if changes are required by the client that are made in the maintenance phase as well [23]. The following considerations should be taken: 1) have all identified vulnerabilities been accounted for, 2) the most efficient to deliver the product, 3) error and vulnerability check during maintenance phase, 4) check for any changes required for mitigation of the errors and improvement in the security aspects [24]. It is crucial to introduce security aspects in all stages of software development process and provide a standard guideline for performing security checks. Although there are analyses of various software development strategies available in literature [25], there is a gap found in terms of investigation on effective security feature integration with the development strategies.

### 2.2 Risk Management

In developing a software application system, risk management plays a vital and special role in the whole project although not everyone knows why risk management is so crucial when it comes to developing a secure software system [26]. Other than having a secure SDLC with GQM approach and also threat modelling, risk management should always be considered in the first place. As a secure software developer, it is required to go through all the process and double-check so that the product produced and presented to the stakeholders and users are reputable. It is also hard to deal with the outcome and need to have a clear explanation to the stakeholders when the results are not really convincing. Risk management is known as a collection of procedures taken by risk managers to oversight the risks that might happen on the system. It is proven that risk management can bring victory to a software development based on previous research [26]. However, sometimes situations can be different where the project leader will neglect to implement risk management due to some of the limitations that they have in mind.

Thus, everyone should raise their awareness to always take risk management into consideration before starting a project. Though a risk manager can be very costly and they require a lot of experience and expertise to determine what are the potential risks, what types of approaches can be used and taking up the courage to make decisions for the project team. Overall, it is safe to prevent than to be regret later, so having a risk manager in a team is beneficial as they do not need to spend extra money to hire an out-sourced risk manager to carry out this analysis when the software system rolled-out to global market and users will complaint about having issues while using the system.

Risk assessment [27] and management is so crucial because it helps improve the efficiency and security of the project tasks being carried out [28]. It is due to the reason that they get to spot all the potential risks that might happen later on and solve it right away. It saves more time and workload for the team members at the same time as all the burdens when a task is not completed on time. Risk management can also improve the communication between one another among the team. This is because while executing decision-making, risk managers will have to keep on communicating and update with the team members to let them know why the risks found were harmful and why they should mitigate it [29]. Not only that, but risk management can also help minimize the unpredictable events. As people say, we must predict the unpredictable like a risk manager will always think a step ahead and predict what possible event can happen in the future and cause the project to be under the risks of failing. With that being said, some of the risk management methods can be implemented are qualitative risk analysis and also quantitative risk analysis. Both of the methods allow risk managers to analyse what are the correlation between the risk occurrence and their impact to the system [29][30].

### 2.3 Threat Modelling

Threat modelling can be done to improve an application's overall security in an economical way. Threat modelling is a way to identify potential vulnerabilities or security flaws inside the design of an application that could be exploited by an attacker. It can be done at any phase of the secure software development life cycle but it is recommended that it should be done in earlier phases preferably design phase as it is easier to make changes early on in the design phase then to change once the application is already been made. Threat modelling also provides certain mitigation techniques to reduce the damage if these vulnerabilities and security flaws were to be exploited by someone. The different levels of risks a vulnerabilities or security flaw poses to the application is also identified in this phase. This makes it easier to manage all the vulnerabilities and security flaws and offer mitigation for them, and as a whole make the application safer. Fig. 3. Shows an example of website threat modelling for reference.

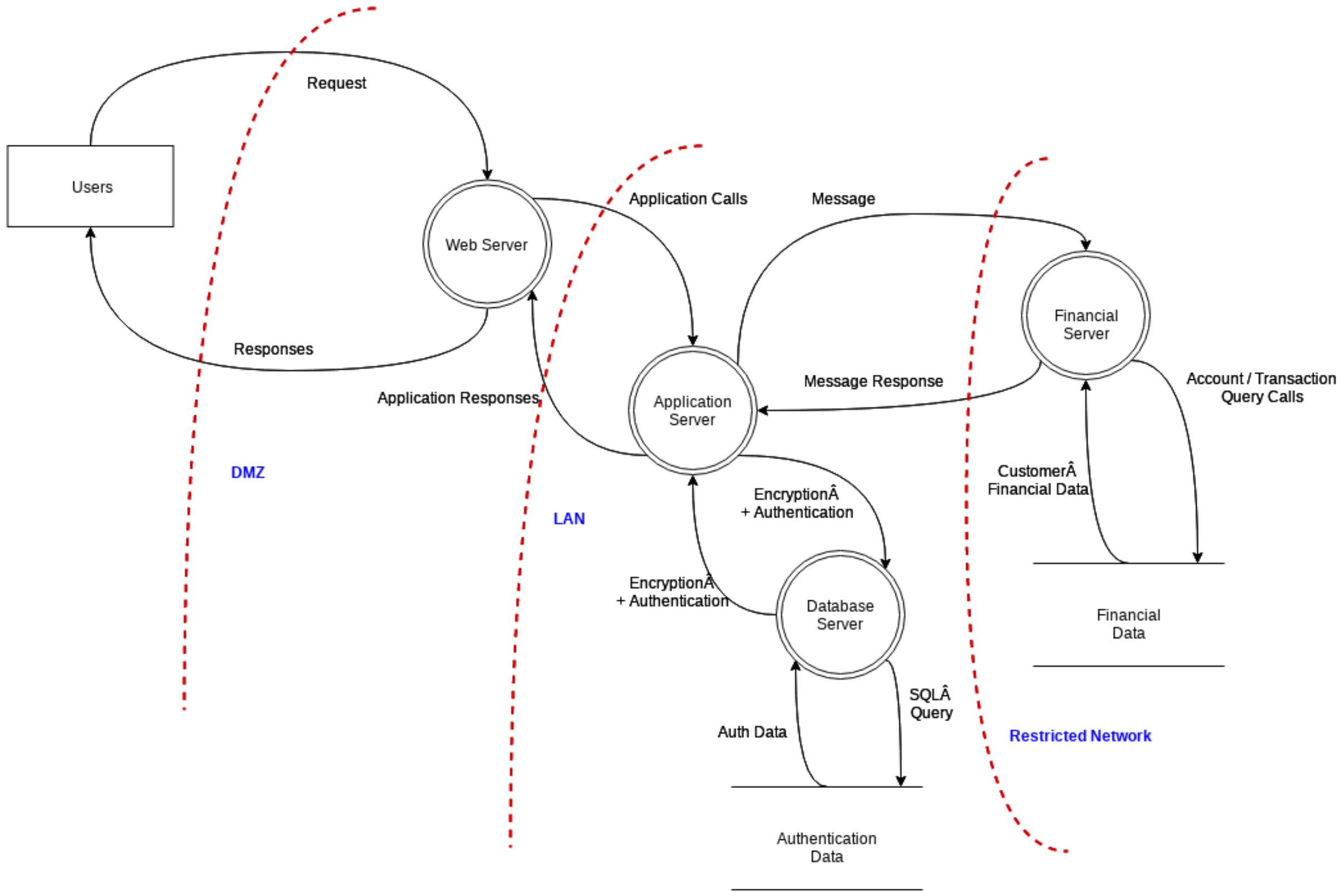

Fig. 3. Website threat modelling [31].

One of the main approaches of threat modelling is software centric threat modelling, where the design of the application is considered. In this type of threat modelling the design of the application is analysed to look for vulnerabilities in the design of the software [32]. To do this it uses various diagrams like data flow diagrams and use case diagrams to have a clearer picture of the design of a specific application and to better identify vulnerabilities in the design of the application. Microsoft's SDL uses software centric threat modelling [33]. Another approach is attacker centric approach, this approach is based on understanding the attacker to make software more secure. The potential attackers of an application are identified, and their various characteristics are identified including their skills and their reason for the attack. This information is used to understand the potential hackers and mitigation techniques can be suggested. Attacker centric approach often use tree diagrams to better understand an attack [34].

Another major approach is Asset centric approach, this approach focuses on the assets of an application for example a database. Characteristics about an asset are identified including how important the data saved in them are and whether the data would be of value to potential hackers [28]. This approach also makes use of attack tree diagrams to identify the different ways an asset can be attacked by an attacker and understand how an asset can be exploited. This understanding is used to suggest mitigation techniques [35]. Threat modelling is a very economical and efficient way to identify various threats to a system, their severity and suggest possible mitigation and improve the overall security of an application. It is specifically important to perform threat modelling when software systems are required for mission critical or sensitive and critical domains like health-related applications [36], nuclear systems, software development for unmanned gadgets [37] and such other areas of applications.

## 3 Methodology

In this section, the methodology and data collection process will be discussed. Reasoning of methodology and information gathering techniques is also justified. Data collection sites and forms are discussed in relevance of gaining information for this research.

### 3.1 Methodologies used and Justification

*A. Scepticism*

From western philosophy, Scepticism is the attitude of doubting knowledge claims set in various areas, which especially used to challenge the reliability and adequacy of the knowledge claims. The claims are challenged by the principles they are based on and what they actually established. Two main types of scepticism are Pyrrhonian scepticism and academic scepticism [38]. In this paper, academic scepticism is incorporated along with demonstration of new information in the research and the validity of the knowledge is challenged to make sure it is reliable and accurate.

*B. Axiology*

From Greek axis means ‘worthy’ or ‘science’, axiology is known as theory of value [39]. Value is meant to make the worth of something and it's used to define the with or something by trading or exchange. The usage of this methodology in this research is to confirm the concept about the value and ethics that are collected for this research.

*C. Ontology*

Ontology is the philosophical study of being in general, or what neutrally apply to everything that is real [40]. An individual that believes in many types of objects will have higher ontology compared to an individual that believes in fewer types of objects. Ontological questions exist in information, which will be used to incorporate this research as a method of gaining information.

### 3.2 Data Collection

In this subsection the sources and sites that are used to gain information regarding the research are briefly discussed. The methods of data collection mainly the published papers, reports and websites.

*A. General Methods*

To get most of the reliable information about the research, the content that is used in this research is limited to articles and publications from the year 2015 or newer. Since there is volatile information regarding the nature of the guidelines for the formation of safe technologies, the limitations are made to make sure updated publications regarding the issue are taken from the year 2015 to 2020. While having reliable information from before 2015 will increase the amount of data for the research, it will risk the provided information to be unreliable and outdated, since the advancement of technologies and guidelines flow changes rapidly as newer implementation occurs. Older information can lead to security issues that may have a newer implementation of the current solution which can create incorrect implementation for the report. Using newer publications can prevent that from happening with statistics being ensured and the publication will be more accurate for the current time being until newer solutions appear. This will also ensure the information provided in this research will be meaningful and correct and does not degrade the quality of the information.

*B. ResearchGate*

ResearchGate was founded in 2008 by physicians Dr. Sören Hofmayer, Dr. Ijad Madisch [41] and computer scientist Horst Fickenscher to share research papers, find collaboration, ask and answer questions for scientists and researchers. It has a large collection of research papers which are published openly or as requested. Content of the research papers are patents, presentation, codes, and many more. By having a research gate, we have an advantage of reliable and reputable sources to be included in the research.

*C. ScienceDirect*

Science Direct was founded in 1997 by Elsevier which is Netherland based information and analytics company. It has 16 million articles in the databases with 39 thousand e-books available for access [42]. The main objective is to be the leading peer-reviewed scholarly research to help researchers, students and more to improve information sharing of the scholarly research. By having access to ScienceDirect, we can find articles and research that can help to produce our research with better quality and reliable information.

## 4 Discussion on Findings

### *A. Usage of GQM*

Goal/Question/Metric (GQM) gives developers more clues on finding the threats and risks that can relate towards the effectiveness of the program that is being developed. Since the GQM method can create different objects and variables, the usage of such techniques can be implemented into Machine Learning to gain multiple results. GQM models can be used to design scenario questions for different patterns and it can be set by natural language processing and Machine learning techniques. Proper process patterns can result in a high precision for a certain project [43]. GQM can also be used in cloud server environments with security metrics derived from cloud storage security key indicators with related security goals. Metrics that have been created can guide the organization in accessing cloud security, especially control security [17]. With the implementation of GQM in development of software, a drop of 43% to 24% errors and bugs can be seen before and after implementation. With the use of indicators in GQM, there has been increased quality of the software development [44].

### *B. Secure SDLC*

Secure Software development Life Cycle (SDLC) is used to create a proper pathway on creating a software with implementation of security in mind. Incidents on software developments usually occur towards the end of the development process. The result of these incidents would be too costly for the company and stakeholders to bear. With the usage of secure SDLC model, it will promote the idea of implementing security measures early in the development process and can be used in any software engineering projects regardless of approach used. With the usage of the model, additional research can be done which will enhance the framework in the software engineering field [22].

### *C. Risk Management*

Risk management is an ongoing process and research found out that no standard method can be applied to all the projects as every project differs from each other [16]. Complexity of risk management increases as the project complexity increases. Tools have been created to assist in risk assessment and found out it has covered any development methodology, either agile or traditional or both are mostly supported. New tools are regularly updated, and such tools are used based on the project needs [45].

### *D. Threat Modelling*

Threat modelling tools are faster in completing full threat analysis rather than manually to be done [46]. Threat modelling helps IT personnel with no previous experience in threat modelling to produce improvements in threat analysis. It also helps large enterprises to use existing resources more effectively to mitigate security threats [47]. With threat modelling, choosing the best modelling method requires specific areas to target which is either risk, security or privacy [48].

## 5 Proposed Solution

In this section, the authors have proposed a unique solution to employ security features at different levels of the development phases. The three approaches GQM, secure SDLC and risk management are combined which are threat modelling software centric, attacker centric and asset centric in nature, respectively. It is observed that the security is mainly applied in design phases generally, but for improvement purposes, the implementation of security features is proposed to be applied in all stages.

To start off all the assets of the application that need to be protected are identified. The main focus is on assets containing sensitive data or resources that are vital to the application and need the most protection. After this the attacker centric approach is used and identification of the possible attackers who might attack our application is performed. The characteristics of a potential attacker are identified and analysis of the potential attackers is performed using attack trees and trees diagrams, which help in the identification process. Then the software centric model is activated and the design phase is analysed, using various diagrams including data flow diagrams and use case diagrams. They are used to identify the possible paths the attacker might take to attack the assets. With all this information the potential vulnerabilities in the application are identified, using the data regarding the assets, the characteristics of the attackers and the design of the application..

Thereafter, risk assessment is performed to identify the most crucial vulnerabilities that need most attention and mitigation techniques for all the vulnerabilities are suggested according to the characteristics of the specific vulnerabilities. When the goals are identified in the initial design phase, GQM can be incorporated as well, and later on during the threat modelling and risk management stages the questions from GQM can be used to ensure that the goals are being met. Once the application is complete, the merit list is used to ensure that the final application is secure and meets the security requirements. The focus is on incorporating the aforementioned steps in the design phase of the secure SDLC to ensure the application is secure when it is developed and major changes are not required to be made later on. The aforementioned steps are aimed to be incorporated in all other phases of as well, for example, in the testing phase these steps can be used to ensure that the finished application is secure. The conceptual workflow of the proposed solution is demonstrated in Fig. 4.

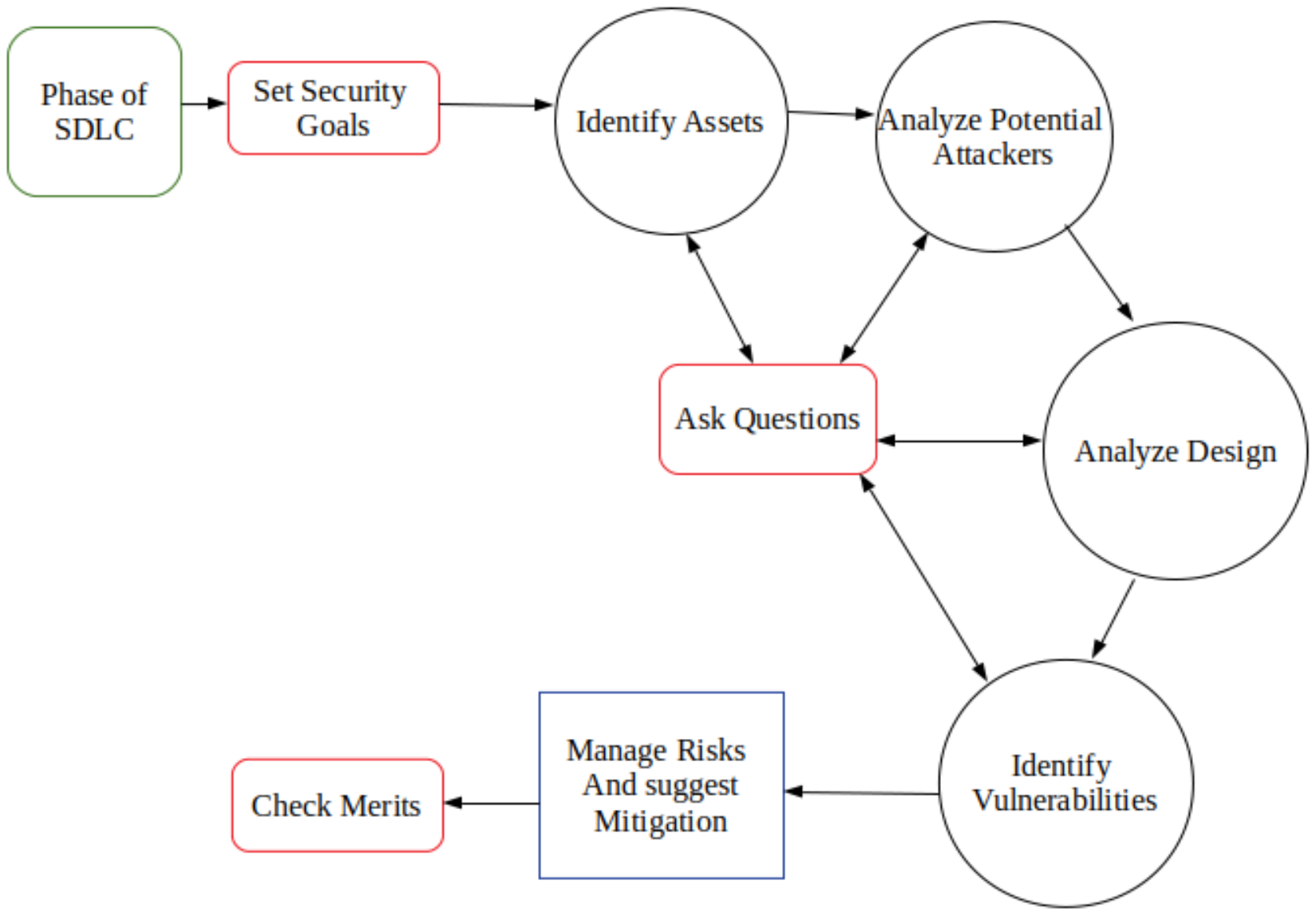

Fig. 4. Workflow of the proposed solution.

## 6 Conclusion

In this paper, the importance of security at different levels of software development is explored and lack of management at higher level is pointed out. Organizations often take security as an afterthought, usually an application will be made without incorporating any security in the SDLC and later on the application will be exploited. This causes tremendous amount of damage to the company reputation-wise and financially as well. Additionally many components or the whole application is required to be reworked upon to address the security issues which is an expensive task. In this paper, three approaches are proposed to be incorporated for improving security aspect in software development process. Incorporating security in all phases of the SDLC can effectively reduce the chances of an attack. For instance, threat modelling helps to identify all the threats in a system before it is launched and offers mitigation solutions so that they are not exploited by an attacker when deployed in real environment. Furthermore, risk assessment and management helps to manage all the different risks present in an application and the GQM helps in identification of the goals of an application and ask questions to ensure the security goals have been met before an application is launched. It is conclude that if appropriate security features are employed in all stages of software development lifecycle, it will save organizations the financial loss that they otherwise have to bear when they fall victim to security attacks.